\documentclass[fleqn,twoside]{article}
\usepackage{espcrc2}
\usepackage[dvips]{graphicx}

\graphicspath{{/home/zs/tex/lat2002/}{/home/zs/tex/slides/lat2002/}}
\DeclareGraphicsExtensions{ps,eps}
\newcommand{\alice}{\textsl{ALiCE}\ }
\newcommand{\be}{\[}
\newcommand{\ee}{\]}
\newcommand{\disable}[1]{}

\title{Improved performance of QCD code on \alice}
\author{Z. Sroczynski\address{
Theoretical Physics, Wuppertal  University, Gau{\ss}stra{\ss}e 20,
D-42097 Wuppertal, Germany}}

\begin{document}

\begin{abstract}
We present results for the  performance of QCD code on \alice, the
Alpha-Linux Cluster Engine at Wuppertal. We describe the techniques
employed to optimise the code, including the metaprogramming of
assembler kernels, the effects of data layout and an investigation
into the overheads incurred by the communication.
\end{abstract}

\maketitle

\section{Introduction}
In a typical lattice QCD project the total run-time of code on a
supercomputing platform is often measured in months or even
years. This means that even a modest improvement in the performance of
the code can yield very tangible benefits. There are two aspects to the
optimisation of code for parallel machines: single-node
optimisation and the minimisation of the overhead incurred by
inter-node communications. 

The former requires that the code be written to take full advantage of the
high performance available from todays advanced hardware,
The latter is of particular importance on cluster
machines, like \alice, where the scalability of code can be a serious
problem. 

\section{Single node optimisation}

\begin{figure}
\begin{center} 
\includegraphics[scale=.49,trim=0 -10 0 0]{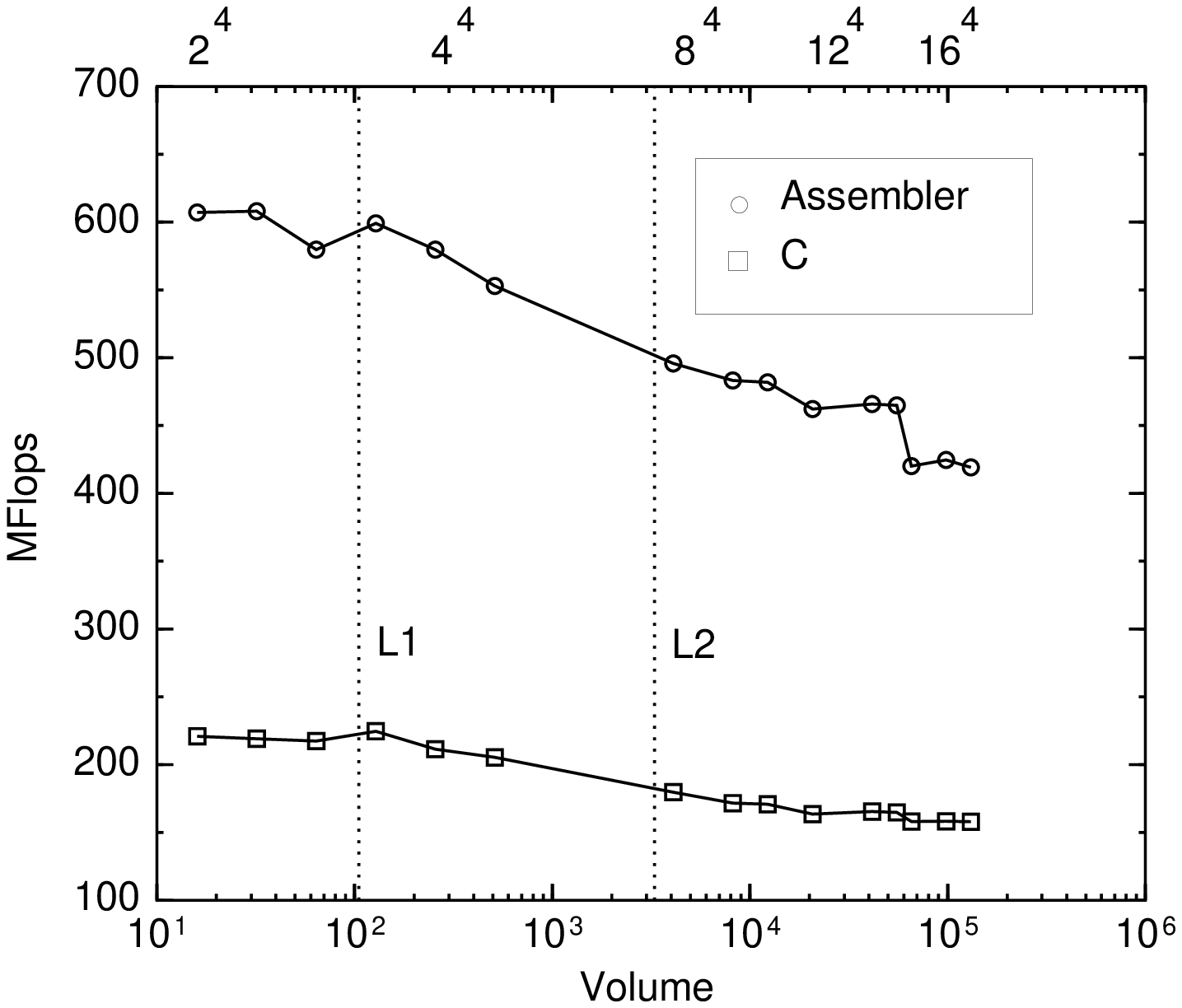}
\includegraphics[scale=.49,trim=0 70 0 -10]{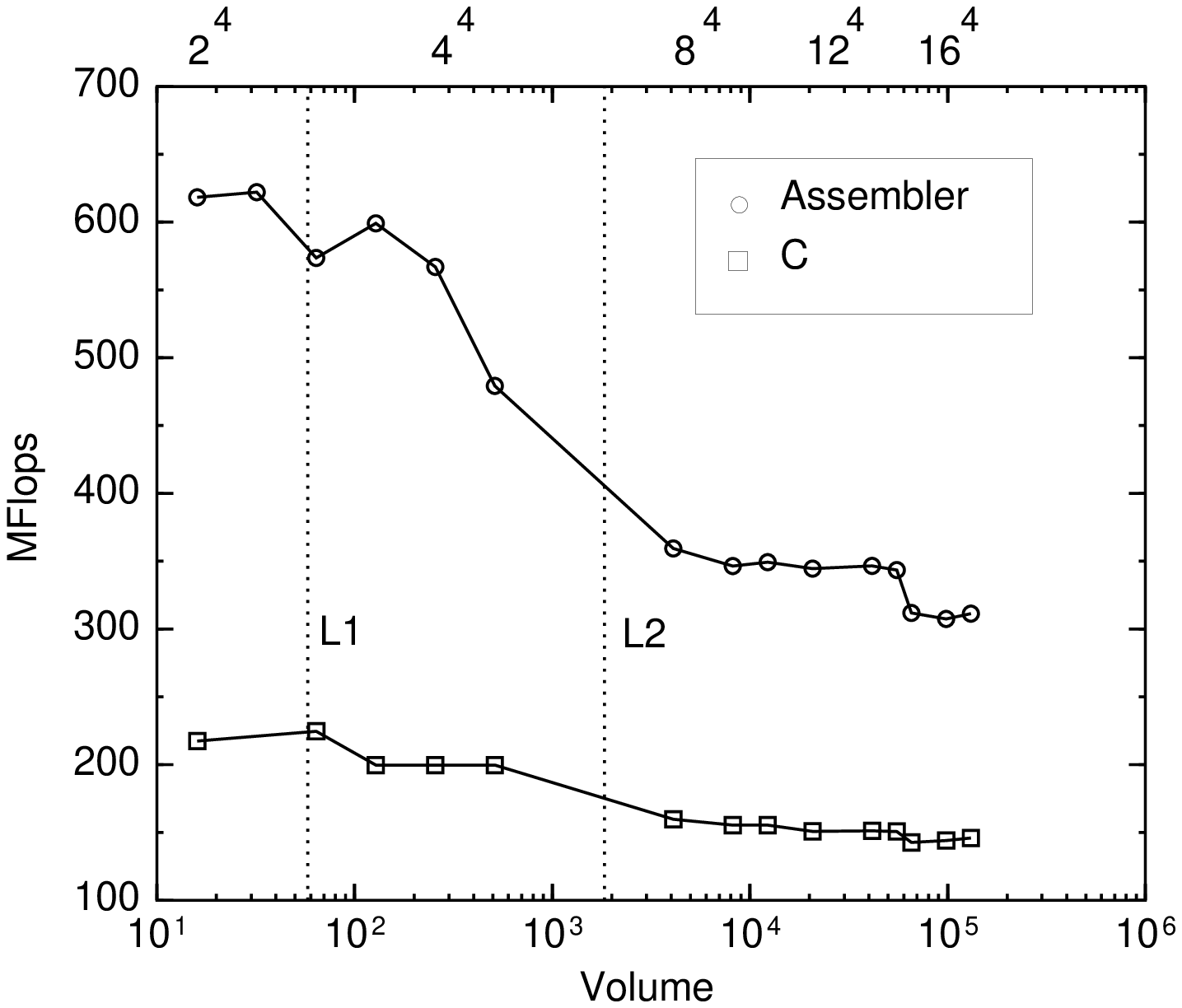}
\end{center} 
\caption{\label{fmm}Wilson matrix multiplication in single (above) and
double (below) precision. The vertical lines indicate the volumes at
which the data fills the level 1 (L1) and level 2 (L2) caches.}
\end{figure}

Experience tells us that the dominant part of a
typical lattice QCD code is that implementing the multiplication of a
vector by the fermion matrix so it is here that the effort
should be made. Secondly, the use of hand-coded optimised assembler
routines can dramatically improve performance since the programmer
can use information about the code which is unavailable to  the compiler.

The disadvantage with assembler routines 
is that they are difficult to develop and harder to maintain, in
addition to the 
obvious lack of portability. We address these problems by adopting a
metacoding approach; writing a C++ program to write the
assembler code for us. We have developed special software tools to
enable this.

\subsection{Metacode software toolkit}

The first stage in creating the assembler routine is to reduce the
computational task to elementary assembler-level abstract
instructions, \textit{e.g.} load a datum
from memory into registers, perform arithmetic on the data, 
cache management, textit{etc.}

In order to write the metacode we have developed a system of C++
classes and routines which automatically schedule the
instructions to hide the instruction latencies as much as possible
and automatically manage the register usage.

When the metacode written using these routines is compiled and run,
the abstract instructions with their arguments
are translated into an actual assembly language and written to a file.

By basing the toolkit design on an abstract RISC ISA it should be possible to
produce assembler code for any RISC machine by changing the
architecture-dependent parameters. Here we show the results on \alice, 
a cluster of Compaq DS10 servers which have a 616 Mhz, 4-way
superscalar Alpha 21264
processor with a 64Kb  2-way set-associative level 1
(on-chip) data cache and a 2Mb level 2 (off-chip) cache.

\subsection{QCD kernels}
An advantage of the metacode toolkit as that it permits a large
degree of flexibility in 
writing various assembler kernels; different approaches can be tried
and compared, and the kernels can be rewritten to adapt to changes in
the action or algorithm.

\begin{figure}
\begin{center} 
\includegraphics[scale=.47,trim=0 70 0 0]{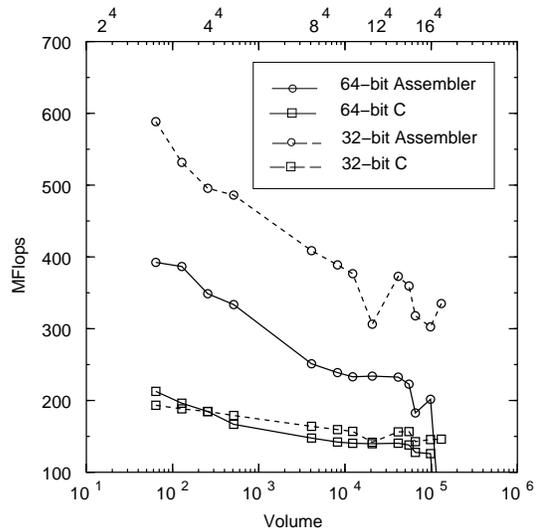}
\end{center} 
\caption{\label{solver}Wilson matrix BiCGstab, comparing single (dashed) and
double (solid)  precision.}
\end{figure}

Figure \ref{fmm} shows the improvement, over a wide range of lattice
volumes, in the performance of the 
Wilson matrix multiplication routine when written with assembler
kernels over that of the original implementation in C. To demonstrate
the effect in a more realistic environment, 
the inversion of the Wilson matrix using BiCGStab is shown in figure
\ref{solver}.

\section{Cluster performance}

\alice is  clustered using ParaStation 3 over 64bit/33MHz Myrinet.
Our code uses  MPICH 1.2.3 to do the communications.

We test the multinode performance of the BiCGstab solver on a
$16^4$ lattice
running on $n$ = 1, 2, 4, 8 and 16 nodes arranged in a
1-dimensional ($1\times n$) grid and a 2-dimensional (square) grid.
We use a standard metric of parallel performance:
\be
\mbox{speedup} = \frac{\mbox{speed on \textit{n} nodes}}{\mbox{speed on 1 node}}
\ee

\begin{figure}
\begin{center} 
\includegraphics[scale=.47,trim=0 70 0 0]{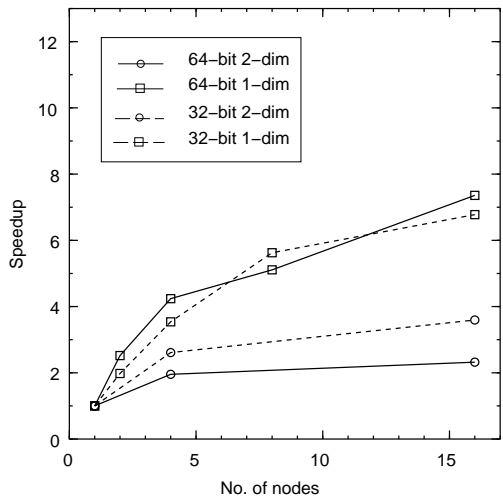}
\end{center} 
\caption{\label{oldspeedup}Speedup of the BiCGstab solver with the
original data layout.}
\end{figure}

\begin{figure}[b]
\setlength{\unitlength}{1in}
\begin{picture}(3,1.2)
\put(.1, 0){\begin{picture}(2, 3.2)
    \put(0,0){
	\includegraphics[scale=.37,trim=0 50 0 10]{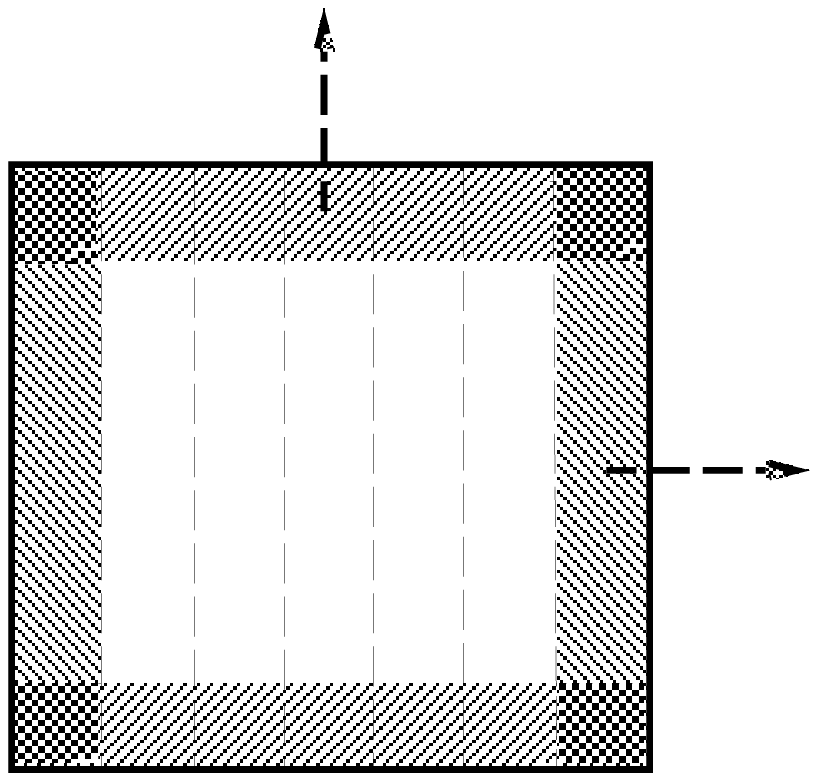}
	}	
  \end{picture}}
\put(1.5,0){\begin{picture}(2, 3.2)
    \put(0,0){
	\includegraphics[scale=.37,trim=0 50 0 10]{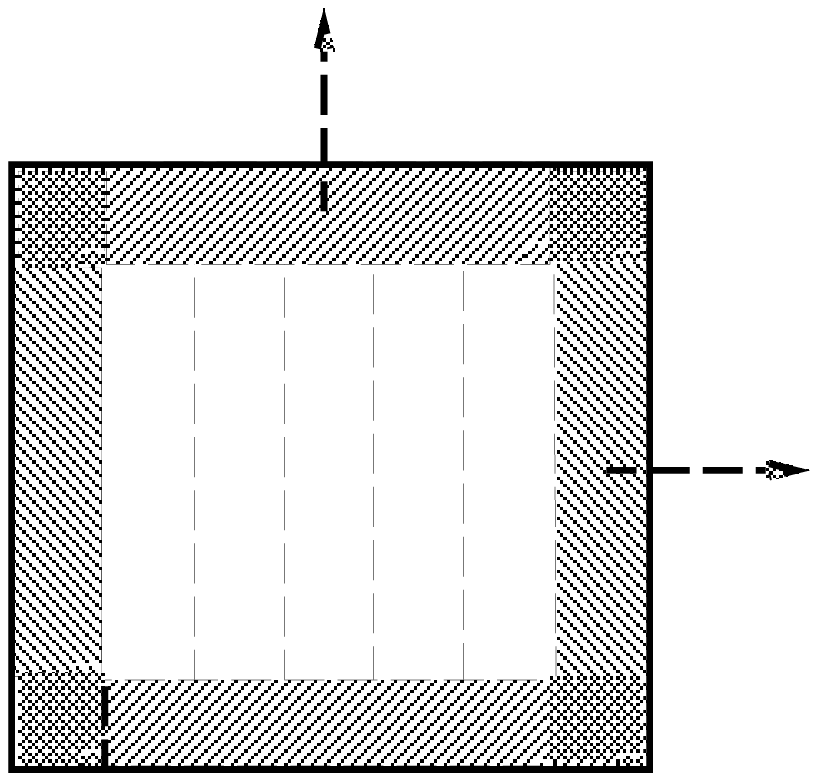}
	}
  \end{picture}}
\end{picture}

\caption{\label{layout} Illustration of the old (left) and new (right)
data layout; the shaded areas show data on the boundary which is communicated.}
\end{figure}

Our original implementation used a conventional array ordering for all
the fields, where each lattice site with coordinates $(x_0,x_1,x_2,x_3)$ is
numbered $n = x_3+N_3x_2+N_3N_2x_1+N_3N_2N_1x_0$ where $N_\mu$
is the size of the local lattice in directon $\mu$. This is
illustrated in figure \ref{layout} (left) which shows that while the
data along the 
boundary in one direction is contiguous, in the second direction it is
strided. Investigations into the performance of our MPI communications
suggest that the communication of strided data introduces an overhead
of at least 20\% compared to contiguous data. This explains the poor
scaling of the solver on a 2-dimensional grid shown in figure 
\ref{oldspeedup}. The scaling on the 1-dimensional grid suffers from
the increasingly unfavourable surface-to-volume ratio of the local lattice.

The solution to these problems appears to be to rearrange the data
layout so that the sites on the lattice boundaries are ordered in a
contiguous fashion, illustrated in figure \ref{layout} (right).

\begin{figure}
\begin{center} 
\includegraphics[scale=.47,trim=0 70 0 0]{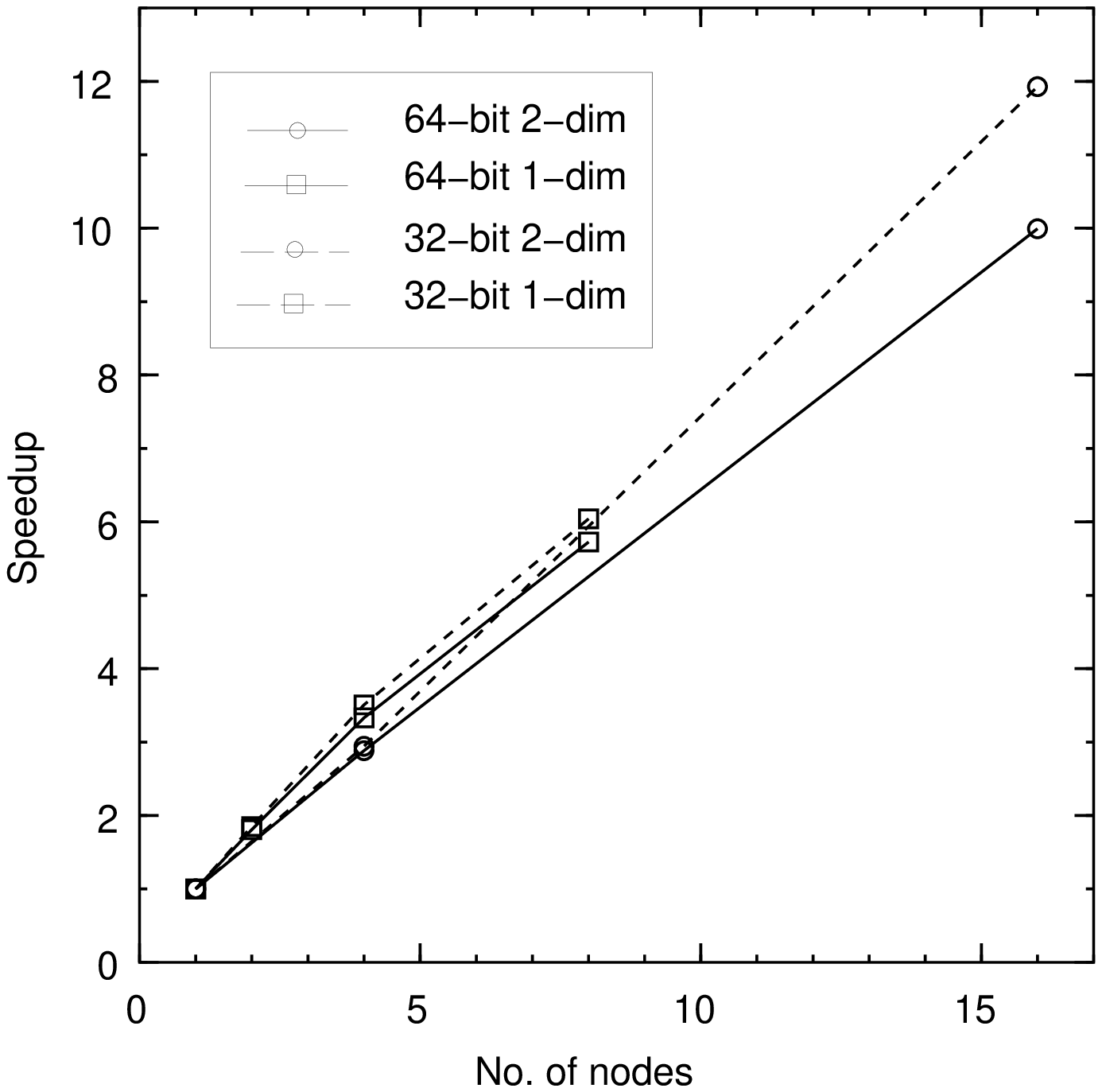}
\end{center} 
\caption{\label{newspeedup}Speedup of the BiCGstab solver with the new
datat layout.}
\end{figure}

Separating the boundary and interior sites in this way has the
additional advantage that computation can proceed on the interior
sites while the boundary sites are waiting for a non-blocking
communication to finish.  
Figure \ref{newspeedup} shows that using this new data
layout greatly improves the speedup of the solver.
The new data layout does not adversely affect single node performance.

\section{Summary}

We have introduced a flexible software toolkit [1] which can successfully generate 
optimised assembler routines for performance-critical parts of our
lattice QCD code. On a single node we see a 100--150\% improvement in
the Wilson matrix solver performance at single precision and 50--100\%
at double precision. 

We demonstrate that good scaling performance can be achieved on \alice
if the data layout and communication strategy is carefully adapted to
suit the communication needs. 

\section*{Acknowledgements}

Thanks to Dr. P. Boyle for inspiration and information and to the
\alice Team.

I acknowledge the financial support provided through the
European Community's Human Potential Programme under contract
HPRN-CT-2000-00145, Hadrons/Lattice QCD.

\section*{References}
\noindent1. \texttt{
www.theorie.physik.uni-wuppertal.de/\\Computerlabor/Alice/akmt.phtml
}
\end{document}